\documentclass[aps,pre,superscriptaddress,twocolumn]{revtex4-1} 

\usepackage{graphicx}

\usepackage[utf8]{inputenc}
\usepackage[english]{babel}
\usepackage{amsmath,amssymb,enumerate}
\usepackage{hyperref}

\usepackage{bm}

\begin{document}


\title{Slow time scales in a dense vibrofluidized granular material } 

\author{Andrea Plati}
\affiliation{Dipartimento di Fisica,  Universit\`a di Roma Sapienza, P.le Aldo Moro 2, 00185 Roma, Italy}
\author{Andrea Puglisi}
\affiliation{Istituto dei Sistemi Complessi - CNR and Dipartimento di Fisica, Universit\`a di Roma Sapienza, P.le Aldo Moro 2, 00185, Rome, Italy}

\date{\today}

\begin{abstract}
Modeling collective motion in non-conservative systems, such as
granular materials, is difficult since a general
microscopic-to-macroscopic approach is not available: there is no
Hamiltonian, no known stationary densities in phase space, not a known
small set of relevant variables. Phenomenological coarse-grained
models are a good alternative, provided that one has identified a
few slow observables and collected a sufficient amount of data for
their dynamics. Here we study the case of a vibrofluidized dense
granular material. The experimental study of a tracer, dispersed into
the media, showed the evidence of many time scales: fast ballistic,
intermediate caged, slow superdiffusive, very slow diffusive. A
numerical investigation has demonstrated that tracer's superdiffusion
is related to slow rotating drifts of the granular medium. Here we
offer a deeper insight into the slow scales of the granular medium,
and propose a new phenomenological model for such a ``secular''
dynamics. Based upon the model for the granular medium, we also
introduce a model for the tracer (fast and slow) dynamics, which
consists in a stochastic system of equations for three coupled
variables, and is therefore more refined and successful than previous
models.
\end{abstract}

\maketitle


\section{Introduction}

Granular materials stand as prototypes of physical systems with both
important industrial applications and fundamental theoretical
challenges~\cite{JNB96b,poeschel}. When the external perturbation is
absent or very weak (and/or sparse in time), a granular medium behaves
as an athermal amorphous solid: in this regime, theoretical approaches
are scarce and can rarely be compared quantitatively with
experiments~\cite{andreotti13,baule18}. The situation is different in
the case of so-called vibrofluidized regime, i.e. when a continuous
external perturbation is applied, for instance by means of vibrating
the box that contains the grains, provided that maximum shaking
accelerations is much larger than gravity
acceleration~\cite{C90,puglio15}. Actually such a regime can be
separated into several different phases, depending upon the amount of
grains in the container (typically measured through the number of
layers at rest, or the average density/packing fraction) and the
shaking parameters~\cite{lohse07,PGVP16}. In the most dilute and
agitated phase, the so-called granular gas, quantitative predictions
are obtained through granular kinetic theory~\cite{poeschel} and
granular hydrodynamics~\cite{BDKS98,sotobook}. Those are bottom-up
theories where macroscopic/average quantities, such as the transport
coefficients, can be deduced from the knowledge of the laws of
interaction among the single grains, with the assumption (true in the
dilute Grad-Boltzmann limit) of molecular chaos, or its revised
version called Enskog
approximation~\cite{poeschel,cercignani,BDKS98,garzo99b,DG01}. In the
last decades, kinetic theory of molecular systems has made important
progresses towards the quantitative understanding of the liquid
phase~\cite{hansen86,EM90}, with success in explaining certain aspects
of slow relaxations in supercooled
liquids~\cite{cavagna09,gotzebook}. A similar approach has been
applied to vibrofluidized granular systems in order to obtain some
predictions in non-dilute
phases~\cite{kranz2013glass,kranz2018rheology}. This approach - based
upon a granular adaptation of mode coupling theory
(MCT)~\cite{gotzebook} - reproduces the qualitative behavior of the
relaxation of density correlations in the system, with the possibility
of marking a glass transition where relaxation times diverge: at the
qualitative level everything appears similar to the molecular
(non-dissipative) case. The diffusional properties of a tracer are
also qualitatively similar to molecular liquids, with the standard
ballistic $\to$ arrested $\to$ diffusive scenario for the mean squared
displacement. Rheological properties and the typical
thinning-thickening scenario are also fairly explained with this
approach, even if experiments may offer more complex
pictures~\cite{gnoli16}.

\begin{figure}
  \includegraphics[width=0.8\columnwidth]{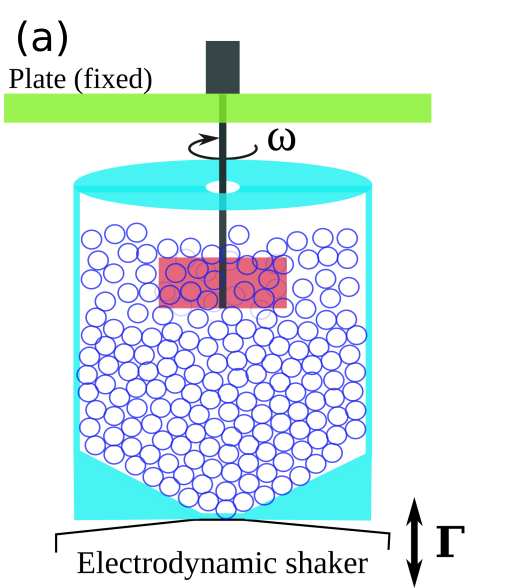}
  \includegraphics[width=0.8\columnwidth]{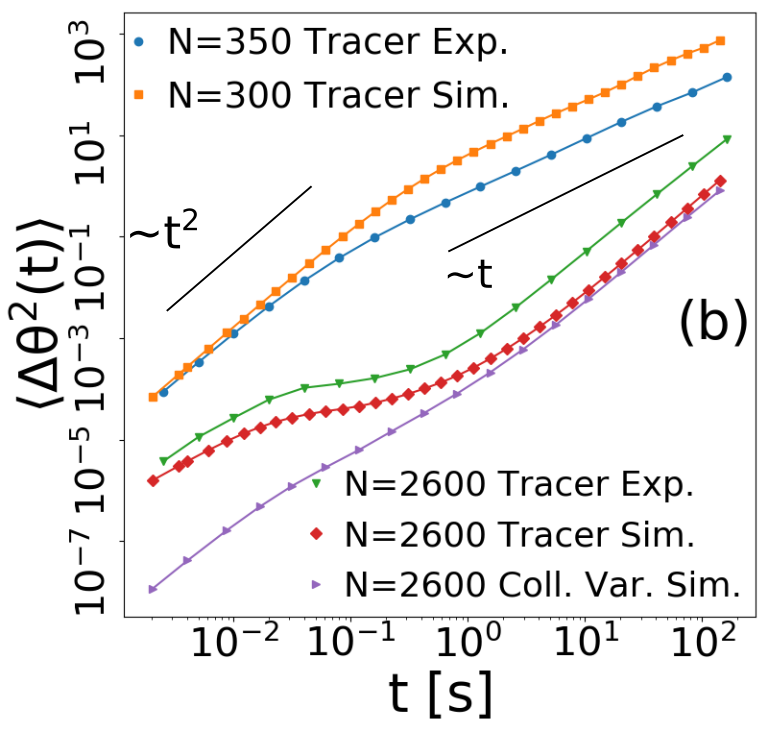}
  \caption{A: Setup of the experiment and of the simulation. B: mean
    squared displacements of the tracer (in experiments and
    simulations) and of the collective rotation of the granular medium (only
    simulations, see Section IV for the definition). $\Gamma=39.8$ for all the cases shown. \label{fig1}}
\end{figure}

Experiments in dense vibrofluidized granular liquids in a cylindrical
geometry (see Fig.~\ref{fig1}) have shown the existence of a complex
dynamical scenario with collective phenomena occurring along
time-scales larger than the cage
time-scale~\cite{scalliet,plati2019dynamical}. The cylindrical
geometry enhances the possibility of observing such time-scales: a
cylinder implies the existence of a direction of motion - the
rotational one - with an infinite horizon, i.e. without obstacles
(only subject to friction).  The slow collective dynamics determines
two observable phenomena: the superdiffusion (SD) for the dynamics of
a tracer immersed in the granular medium~\cite{scalliet}, and
persistent rotational motion (PR) of large parts of the granular
medium~\cite{plati2019dynamical}: both phenomena - strictly correlated
- appear when the density is increased and/or the steady
vibro-fluidization is reduced, and take place over timescales in the
range of $10 \div 10^3$ seconds, depending upon packing fraction and
shaking parameters, while interactions occur on time-scales of
$10^{-3} \div 10^{-2}$ seconds. Interestingly these unprecedented SD
and PR are superimposed to the usually observed fast phenomena
occurring over small and medium timescales, such as sub-collisional
ballistic motion and transient dynamical arrest due to caging. Such
kind of persistent motion is not accounted for by the aforementioned
granular mode coupling theory, since velocities are factored out in
that theory.

The rotational experiments are not unique - in the context of granular
materials - to show some kind of collective motion occurring over very
long time-scales. A long history of slow convective phenomena exists
in regimes where the external vibration is exceedingly weak (below the
gravity threshold)~\cite{ehrichs1995granular,mobius2001size}, but we
consider those cases substantially different from the case considered
here, since they display basically {\em only} long time-scales (no
fast phenomena are present). More recently, a series of
vibro-fluidized experiments has revealed the so-called ``low frequency
oscillations'' LFO ~\cite{rivas2013low,folli13}, with frequencies
below $1$ Hz, superimposed to standard fast relaxational behavior. LFO
and SD/PR phenomena occur over different ranges of frequency (LFO
close to $1$ Hz, SD/PR close to $0.1$ Hz) and their connection is
still to be analyzed.

It is also interesting to compare the slow collective motion seen in
these granular experiments with similar phenomena observed in models
and experiments with active matter, for instance in dense populations
of bacteria~\cite{rabani2013collective,chen2017weak} and
sperms~\cite{tung2017fluid}, or even within the dynamics of a single
flagellum~\cite{battle2016broken}.

In this paper, with the help of discrete-element simulations
reproducing the original experiment in~\cite{scalliet}, we aim at
elucidating some phenomenological stochastic models with few
coarse-grained variables which are able to describe the SD and PR
phenomena~\cite{baldovin18}. These models are an extension and
improvement of a previous one proposed in~\cite{lasanta2015itinerant}
which was built to reproduce only the SD tracer behavior: limitations
of that model are that it is not entirely coherent with the observed
tracer power spectra and does not describe the granular medium
(i.e. it cannot account for PR). These models share minimality (in the
spirit of Occam razor), in fact they include only linear couplings and
additive independent noises, allowing for an exact analytical
treatment: their goal is to characterize the existence of many
timescales in the system and for this purpose non-linearity is not a
crucial ingredient. A more general (not limited to linear coupling)
Langevin approach has been also considered by applying a Langevin
modeling recipe to the experimental data~\cite{baldovin19}. Linearly
coupled models have been used before in granular modeling: for
instance they are able to characterize the failure of
fluctuation-dissipation relation for the dynamics of a tracer in
certain models~\cite{villamaina09,sarra10b}, but have shown certain
limitations when compared with experiments~\cite{GPSV14}. Linear
modeling used for the purpose of quantifying non-equilibrium features
have been also used, recently, in the single-flagellum dynamics of
active particles~\cite{battle2016broken}.

The manuscript is organized as follows. In Section II we give a brief
account of the existing stochastic models for this particular granular
setup. In Section III the ingredients of the simulations are
explained. In Section IV we show how the collective rotation of the
granular medium is well reproduced by the superposition of two {\em
  independent} collective variables obeying linear Langevin equations
with well-separated timescales. In section V we build upon the
previous observation and describe the motion of a rotating tracer as a
third rotating variable coupled to the granular medium, resulting {\em
  de facto} in a three-variable model which shows a close
adherence to experimental and numerical power spectra. In Section VI
we make a more extensive comparison of the two models with
numerical and experimental data in order to rationalise the dependence of the models'
parameters upon the physical parameters of the system. A more general
discussion of the salient features of the proposed models, together
with conclusive perspectives, is given in Section VII.

\section{Existing Langevin models}

A simple and old model of diffusion in dense liquids is the so-called
Itinerant Oscillator model, where the tracer is caged in a (harmonic)
potential well, whose minimum's position is not fixed but slowly
diffuses~\cite{sears65,vollmer79}: this model helped in rationalising
spectra from neutron scattering experiments on
liquids~\cite{rahman64}. If diffusion of the potential minimum is
slower than the particle's diffusion inside the well, then the
behavior of the tracer's mean squared displacement (MSD) shows a
transient plateau (equivalent to the dynamical arrest or caging)
followed by ordinary diffusion. If the dynamics inside the well is
underdamped, the first part of the tracer's MSD time-dependence is
ballistic $\sim t^2$. The after-cage part, however, is always of the ordinary
diffusive type $\sim t$ as the tracer is slaved to the dynamics of the
minimum's position which is purely diffusive. The extension of this model proposed in~\cite{lasanta2015itinerant}
was aimed at obtaining long superdiffusive regimes {\em after} the cage
time. In order to obtain this result, an underdamped dynamics was
considered also for the position of the well's minimum. Such an
extension takes the form of four coupled equations for the angular
velocity and position of the rotating tracer, $\omega(t)$ and
$\theta(t)$ respectively, and for the angular velocity and position of
a collective slow variable which  represents the effect
of a large group of particles surrounding the tracer (the cage),
$\omega_0(t)$ and $\theta_0(t)$ respectively:
\begin{subequations} \label{lsmodel}
\begin{align}  
I \dot{\omega}(t)&=-\gamma \omega(t) -k[ \theta(t) - \theta_{0}(t)]+ \sqrt{2\gamma T}\eta(t)\\ 
I_{0}\dot{\omega}_{0}(t)&=-\gamma_{0} \omega_{0}(t) +k[ \theta(t) - \theta_{0}(t)]+ \sqrt{2\gamma_0 T_0}\eta_{0}(t) \label{lsmodel2} \\
\theta(t)&=\int^{t}_{0}\omega(t')dt' \label{lsmodel34} \;\;\;\;\;\;\;\;\;\; \theta_{0}(t)=\int^{t}_{0}\omega_{0}(t')dt'. 
\end{align}
\end{subequations}
In the above equations $\eta(t)$ and $\eta_{0}(t)$ are independent Gaussian
white noises with zero average and unitary variance, namely
$\langle \eta(t) \eta(t') \rangle= \delta (t-t')$ and $\langle
\eta_{0}(t) \eta(t') \rangle= \delta (t-t')$. The parameters $I$ and
$I_0$ are the inertia of the tracer and of the surrounding medium,
$\gamma$ and $\gamma_0$ are the dissipation felt by the two variables,
and $T$ and $T_0$ are ``temperatures'' (assuming unitary Boltzmann
constant $k_B=1$), see~\cite{sarra10b} for a discussion of their physical
interpretation. The coupling between the tracer and the collective
variable, in this model, is represented by the term $-k[ \theta(t) -
  \theta_{0}(t)]$ which is linear in the {\em positions}. With large
enough values of $I_0$, the inertia of the cage, the model - which is
analytically solvable - reproduces long time ballistic superdiffusion.

\begin{figure}
  \includegraphics[width=\columnwidth]{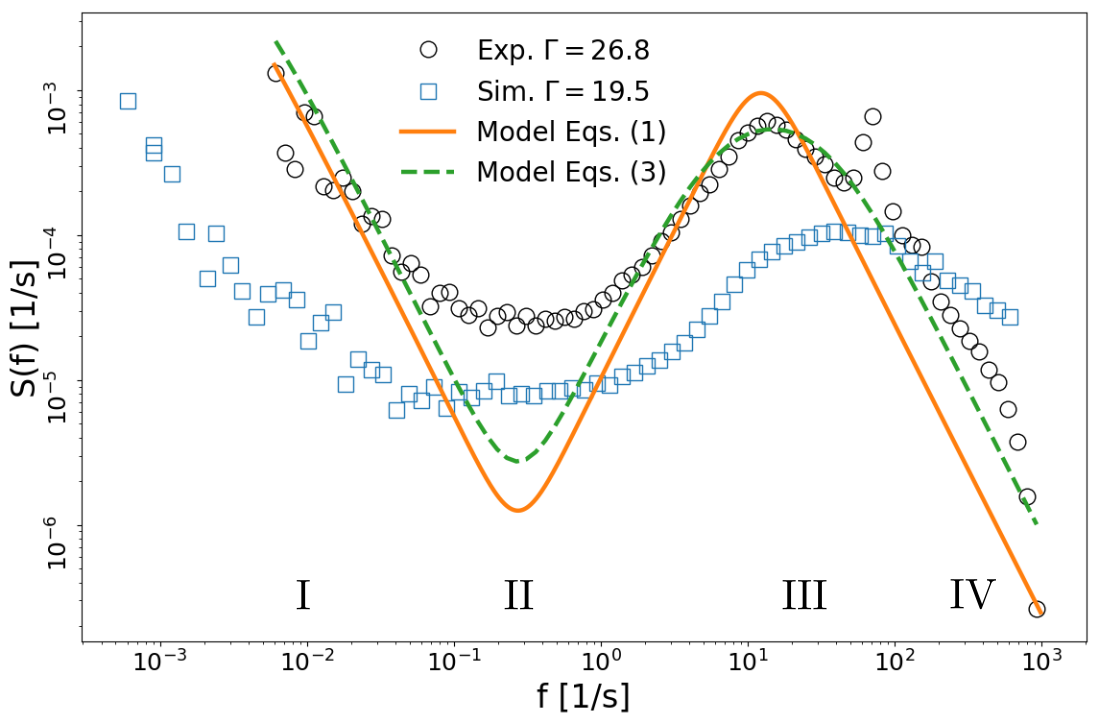}
  \caption{Comparison of the tracer's velocity power spectra (PDS) in
    experiments/simulations and the ones predicted (just for the experiments) by  Eqs.~\eqref{lsmodel} and \eqref{model5}. The two models fail in reproducing the data in region (II) in the same way. Both experiments and simulations are performed with $N=2600$. The peak close to $100$ Hz seen in the experimental spectrum is directly related to the driving frequency and internal mechanical resonances of the setup, see~\cite{scalliet} for details. \label{fig2}}
\end{figure}

A problem of this model however is its inability to entirely reproduce
the velocity power density spectrum (PDS) which is put in evidence in
Fig.~\ref{fig2}. We recall the definition used here for velocity power spectrum
\begin{equation}
S(f)=\lim_{t_{TOT}   \rightarrow \infty} \frac{1}{2\pi t_{TOT}}\left|\int^{t_{TOT}}_{0}\omega(t)e^{2\pi fi}dt\right|^{2},
\end{equation} (where $t_{TOT}$ is the total simulation time)
which is also equivalent to the Fourier transform of the
auto-correlation function in the steady state
$\langle\omega(t)\omega(0)\rangle$. The experimental/numerical power
spectrum - read from high frequencies to low frequencies - shows four
main regions (labels are marked in Fig.~\ref{fig2}): IV) at very high
frequency a power-law decay (slower than the pure Lorentzian case
$\sim f^{-2}$ expected for exponentially decaying velocity
autocorrelations), III) a bump-like peak at smaller frequencies
representing the almost periodic oscillations inside a cage, II) a
plateau at even smaller frequencies (suggesting a range of timescales
where the process rapidly loses memory) and finally I) at the smallest
observable frequencies (unless day-long experiments are conducted) a
decay analogous to the high-frequency one. This decay at very small
frequencies is in fact interpreted, in the above model, as the
high-frequency decay of a very slow Ornstein-Uhlenbeck process
described by $\omega_0(t)$.  The most evident discrepancy between
theory and experiments/simulations is in the central plateau region
(II): the inertial itinerant oscillator model is not able to reproduce
it.

A second attempt to obtain a meaningful Langevin model for the angular
velocity of the rotating tracer has been done in~\cite{baldovin19}. Its
advantage is that it is the result of a general constructive method of
Markovian model-building based upon the experimental data (the
experimental time series of $\omega(t)$) and some guess about other
possibly relevant variables (when data present non-Markovian
character, other variables must be identified in order to obtain a
proper Markovian embedding~\cite{baldovin18}). In such a method one is also able to
verify that the chosen variables are consistent with the Markovian
hypothesis. Such an advantage results in a more compact model, with a
smaller number of parameters:
 \begin{subequations}
  \label{model5}
   \begin{align}
    \dot{\omega}(t)&= - A_1 \omega(t) -A_2 [\theta(t)-\theta_0(t)]+ \sqrt{2B}\eta(t)\\
    \dot{\omega_0}(t)&=-A_0 \omega_0(t) + \sqrt{2B_0}\eta_0(t)\\
    \theta(t)&=\int^{t}_{0}\omega(t')dt'  \;\;\;\;\;\;\;\;\;\; \theta_{0}(t)=\int^{t}_{0}\omega_{0}(t')dt',
  \end{align}
\end{subequations}
where $\eta(t)$ and $\eta_0(t)$ are Gaussian noises with unitary
variance. This model is a particular limit of the model guessed
in~\cite{lasanta2015itinerant}, where the term
$k[\theta(t)-\theta_0(t)]$ is negligible with respect to the other
terms in Eq.~\eqref{lsmodel}b: indeed with the present definition,
$\omega_0$ evolves with a slow dynamics that does not admit
fluctuations on the fast time scale of $\theta_1$, so that such term
is necessarily negligible. All other parameters are strictly related
to the parameters of the model in Eq.~\ref{lsmodel}: $A_0$, $A_1$,
$A_2$, $B$ and $B_0$ should be directly compared with $\gamma_0/I_0$,
$\gamma/I$, $k/I$, $\gamma T/I^2$ and $\gamma_0 T_0/I_0^2$
respectively. This model well reproduces the mean squared displacement
of the tracer with its final superdiffusive part, while its comparison
with the power spectrum is unsatisfying in the central part, as
in~\cite{lasanta2015itinerant}, see Fig.~\ref{fig2}.

Models in Eqs.~\eqref{lsmodel} and~\eqref{model5} both involve two independent white
noises but they are anyway a threedimensional linear model in which the
Markovian vector is $X(t)=\{z(t),\omega (t),\omega_0(t)\}$ with
$z(t)=\theta(t)-\theta_0(t)$. One may wonder  if it is possible
to obtain satisfying results with a 2-variable linear model. We know
from the data that a minimum request is a PDS with two stationary
points (a minimum and a maximum) for $f>0$. In the appendix we show
that this kind of functional form is not compatible with the most
general form for a two-variable linear model. In view of this, using a
three variable model is an unavoidable choice if one wants to reproduce
the experimental data through stochastic linear models.


\section{Microscopic model}

Simulations of the granular experiments are implemented through the
LAMMPS package with its module dedicated to granular
interactions~\citep{plimpton1995fast}. LAMMPS is an optimized package
that solves the molecular dynamics equations of motion and in our case
incorporates the interactions of discrete elements methods to treat
macroscopic (non molecular) particles such as the spherical grains
of the experimental setup.  Specifically all interactions among the
bodies in the simulation, including interactions with the boundaries
representing the experimental box, obey the Hertz-Mindlin
model~\citep{Zhang2005,Silbert2001,Brillantov1996}: the latter is a
soft potential that takes into account normal and tangential forces,
both made of elastic and dissipative contributions.  Thus, our
particles are spheres of mass $m_i$, radius $R_i$ and momentum of inertia
$I_i=\frac{2}{5}m_iR_i^2$ with spatial coordinates $\vec{r}_i$ that
travel with translational velocities $\vec{v}_i$ and rotational
velocities $\vec{\omega}_i$. We specify that the flat boundaries of
the box are considered as spheres with infinite mass and radius. When
centers of mass of two particles are closer than the sum of their
radii, a collision takes place, which is the only situation in which
the interaction forces are non-zero: in that case particles $i$ and
$j$ compenetrate to each other and a relative velocity at the surface
of contact is defined as $\vec{g}_{ij}=
(\dot{\vec{r}}_{i}-\vec{\omega}_{i}\times
R_{i}\vec{n})-(\dot{\vec{r}}_{j}+\vec{\omega}_{j}\times R_{j}\vec{n})$
where
$\vec{n}=\left(\vec{r}_{i}-\vec{r}_{j}\right)/\left|\vec{r}_{i}-\vec{r}_{j}\right|$;
we call $\vec{g}_{ij}^{N}$ and $\vec{g}_{ij}^{T}$ the two projections,
respectively normal and tangential, to this surface of contact. Now
the equations of motion read as:
\begin{subequations} \label{force}
\begin{align}
\dot{\vec{g}}^{N} =\vec{F}^{N}/m^{\text{eff}} \label{forceA}\\
\dot{\vec{g}}^{T} =\dfrac{7}{2m^{\text{eff}}}\vec{F}^{T} \label{forceB},
\end{align}
\end{subequations}
where $m_{ij}^{\text{eff}}=m_{i}m_{j}/(m_{i}+m_{j})$ while
$\vec{F}_{ij}^{N}=\vec{n}(\vec{n}\cdot \vec{F}_{ij})$ and
$\vec{F}_{ij}^{T}=\vec{F}_{ij}-\vec{F}_{ij}^{N}$ are respectively the
normal and tangential component of the force $\vec{F}_{ij}$ between
the particles. Both these two contributions are made of an elastic and
a dissipative term, tuned by coefficients that depend upon the
properties of the specific modeled material. To be more compact, we
avoid to write here the full form of these terms, reported and
explained in detail in the supplemental material of our previous study
\citep{plati2019dynamical}. In the same reference it is possible to
find also the specific numerical values used to tune our
simulations. The simulation setup fairly mimics the experimental setup
of \citep{scalliet} with all its specific components and materials as
the cylindrical box with the conical base and the blade used as
intruder.  The good quantitative agreement between the experimental
and the numerical observations for the intruder has been shown
in~\citep{plati2019dynamical} (see also Fig. \ref{fig1} here). In the
same paper simulations without the blade have also been performed, in
order to study the collective motion of the granular medium,
inaccessible in the experiments. These studies confirmed that the PR
phenomenon displayed by the granular medium is not affected by the
presence of the tracer. Nevertheless we specify here that, except when
explicitly declared, all the data relative to the collective motion
shown in this paper (see $\Omega$ defined in the next section) come
from simulation performed \emph{without} the blade.  To conclude this
section, we mention that during the present numerical study an error
in the source code of LAMMPS has been
found~\citep{Lammps:mailList}. In particular, the tangential force
during the collision was always applied at the surface of the
particles (i.e. at a distance $R_i$ from the center) and this
naturally lead to an unphysical resultant torque during contact. This
error is actually critical from a physical point of view because it
breaks the conservation of the total angular momentum expected in
internal (sphere-sphere) interactions. Nevertheless, we have corrected
the code and run all the simulations again verifying that this error
does not affect our old and new results in a significant way. Moreover
from the LAMMPS stable release of 5 June 2019 the granular module has
been updated and the error fixed \citep{Lammps:releaseJune}.






\section{Collective variable (two time scales)}
\label{sec:sec4}
The results presented in~\cite{plati2019dynamical} demonstrate that
the SD phenomenon displayed by the rotating tracer is a direct
consequence of the PR phenomenon exhibited by the granular medium:
while on small time-scales the rotating tracer has its own dynamics
with short free flights and rapid bounces against the boundaries of a
local cage of surrounding grains, on long time-scales the tracer is
dragged by a persistent collective rotation of the surrounding
medium. Such a medium rotation is measured through a global angular
velocity defined as
\begin{subequations}
  \begin{align}
\Omega(t)&=\frac{1}{N}\sum_{i=1}^N \dot{\theta}_i(t) \label{eq:OmegaDef1}\\ 
\theta_i(t)=\arctan\left(\frac{y_i(t)}{x_i(t)}\right) &\;\;\;\; \dot{\theta}_i(t)=\frac{ ({\mathbf r}_i(t) \times {\mathbf v}_i(t))_z}{r_i^2},
  \end{align}
\end{subequations}
where $N$ is the number of granular particles, ${\mathbf v}_i$ is the
velocity of particle $i$, and ${\mathbf r}_i$ is the position of
particle $i$ with respect to a coordinate system such that the origin
lies on the axis of rotational symmetry of the setup. The
time-integral of the global angular velocity represents a global
absolute angle
\begin{equation}
\Theta(t)=\int_0^t \Omega(t') dt'.
  \end{equation}
Such a variable, when density is increased and vibro-fluidization is
weakened, exhibits long persistent drifts in both clockwise and
anti-clockwise direction. This implies the appearance of
super-diffusion for its mean-squared angle $\langle \Delta\Theta^2 \rangle(t)=\langle
[\Theta(t'+t)-\Theta(t')]^2 \rangle \sim t^\beta$ with
$\beta>1$, as seen in Fig.~\ref{fig1}. Conversely, the power spectrum
of the global angular velocity $\Omega(t)$ shows a power-law decay
at very low-frequency. We recall that ordinary diffusion ($\beta=1$) at
long times must correspond, in the power spectrum of the velocity, to
a plateau at low frequencies. The low-frequency decay is the symptom
of persistent motion at long time-scales.

In this Section we propose a new model for the dynamics of the
collective granular rotation. The model is based upon the same
principles applied before to the motion of the tracer, i.e. that
persistent memory can be reproduced by considering (at least) an
auxiliary slow variable. An analysis of the data coming from numerical
simulations showed, however, that in this particular case the
modeling is even easier.

The model we propose for the dynamics of $\Omega(t)$ is the sum of
two independent variables, a fast and a slow one. This direct
superposition is different from the model
in~\cite{lasanta2015itinerant} where the tracer's velocity is always
(harmonically) coupled to the slow variable, even at small
time-scales. In that case the slow variable represents cage dynamics
and the coupling describes the natural confining interaction between
the cage and the tracer. However such a cage dynamics is not present
in the collective rotation (no cage exists for the rotational mode of
the whole granular medium) and therefore there is not a simple
mechanism coupling fast and slow motion: at a first level of
approximation we can consider them to be decoupled. We choose for
simplicity two independent Ornstein-Uhlenbeck (OU) processes with two
different characteristic times $\tau_1=I_1/\gamma_1$ and
$\tau_2=I_2/\gamma_2$ and, in general, two different temperatures $T_1$
and $T_2$. In summary the model is described by:

\begin{subequations} \label{eq:casiSlowFast1}
\begin{align}
\Omega(t)=\Omega_1(t)+\Omega_2(t)\\
I_1\dot{\Omega}_1(t)=-\gamma_1\Omega_1(t)+\sqrt{2T_1\gamma_1}\eta_1(t)\\
I_2\dot{\Omega}_2(t)=-\gamma_2\Omega_2(t)+\sqrt{2T_2\gamma_2}\eta_2(t). 
\end{align}
\end{subequations}
In fact the model can be rewritten with a smaller number of
parameters: the only coefficients that count are $\tau_i$ and
$q_i=T_i/I_i$ with $i=1,2$.

We are in the presence of a sum of two independent variables,
therefore the PDS and the MSD are simply the sum of the two individual
OU contributions:
\begin{subequations} \label{eq:casiSlowFast2}
\begin{align}
\langle \Delta\Theta^2 \rangle(t)= 2 q_1 \tau_1 t+ 2 q_1 \tau_1^2 (e^{-\frac{t}{\tau_1}}-1) \notag \\+2 q_2 \tau_2 t +2 q_2 \tau_2^2(e^{-\frac{t}{\tau_2}}-1)\label{eq:casiSlowFast2msd}\\
S(f)=\frac{q_1 \tau_1}{\pi [1+(2\pi f \tau_1)^2]}+\frac{q_2 \tau_2}{\pi [1+(2\pi f \tau_2)^2]}.\label{eq:casiSlowFast2pds}
\end{align}
\end{subequations}
Our idea is then to consider one of the two characteristic times much
larger than the other ($\tau_2 \gg \tau_1$). To make clear now the
meaning of the two variables, we expect the slow component of the
collective variable, $\Omega_2$, to behave similarly to the filtered
variable
\begin{equation}
  \Omega_s(t)=\frac{1}{\tau_f}\int_{t}^{t+\tau_f}\Omega(t')dt',
\end{equation}
obtained with a moving average of $\Omega(t)$ over a time $\tau_f$ such that $ \tau_2\gg
\tau_f\gg\tau_1$.  In order to verify this conjecture we proceed in two ways: first we try
to fit the numerical MSD and PDS via Eqs. \eqref{eq:casiSlowFast2}, then
we show that the superdiffusive part at late times of the collective
MSD coincides with the MSD of the filtered variable $\langle \Delta
\Theta_s(t)^2\rangle$ where $\Theta_s(t)=\int_{0}^{t}\Omega_s(t')dt'$.

\begin{figure}
  \includegraphics[width=0.8\columnwidth]{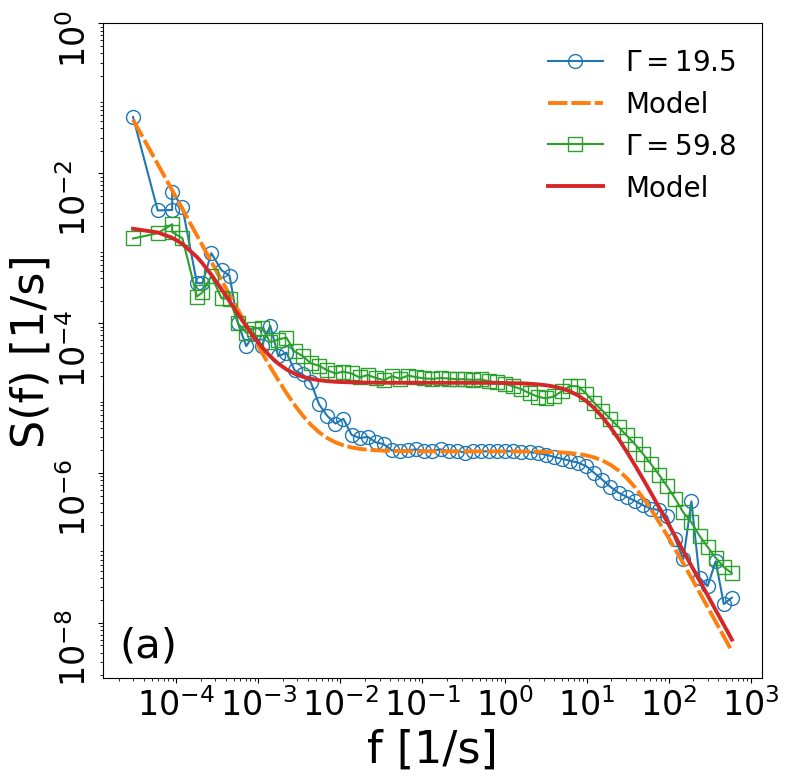}
    \includegraphics[width=0.8\columnwidth]{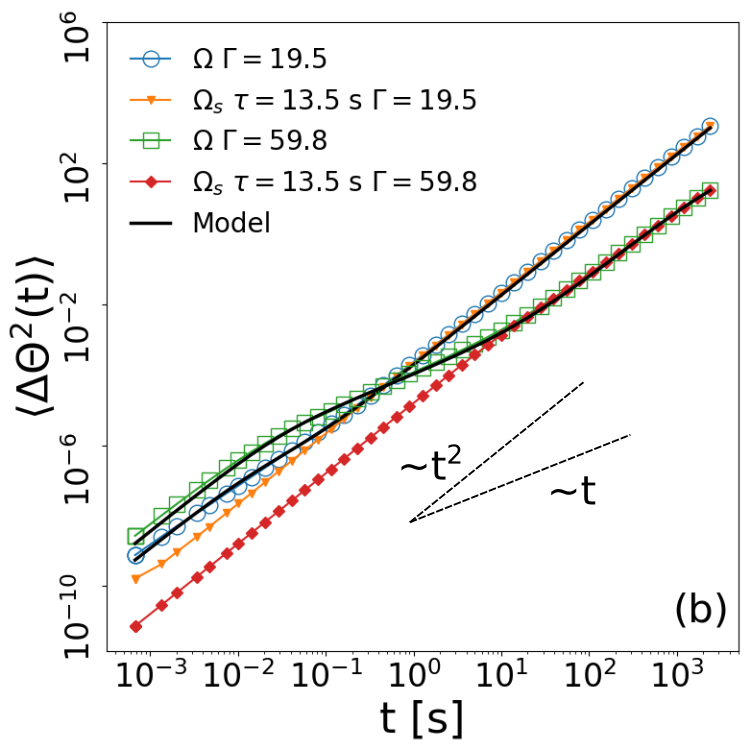}
\caption{Collective motion. First comparison between model predictions
  and numerical data ($N=2600$, $\Gamma=19.5-59.8$ ). MSD and PDS refer
  to the same signal. The simple model in
  Eqs.~\eqref{eq:casiSlowFast1} fits well the numerical data for both
  the PDS and the MSD. The fitted parameters are reported in Table
  I. \label{fig3:firstCompariWithData}}
\end{figure}

In Fig. \ref{fig3:firstCompariWithData} we show how
Eqs. \eqref{eq:casiSlowFast2} can fit the numerical data for two
particular cases of control parameters ($N=2600$, $\Gamma=19.5-59.8$),
postponing a more systematic analysis to section VI.  In order to
obtain the theoretical lines we have first performed a fit of the PDS
via Eq. \eqref{eq:casiSlowFast2}b and then used the parameters
inferred in this way also for the MSD
(Eq. \eqref{eq:casiSlowFast2}a). We can see that the model properly
predicts the behaviour of the two observables. In the PDS there is a
good agreement at all the frequency regimes except for the high
frequency decay where a linear model can only predict a $f^{-2}$
behavior while the data show $f^{-\alpha}$ with $2 >\alpha >1$. The
MSD also exhibits an almost perfect agreement between data and model
predictions at all time scales. We conclude that the idea of
decomposing the total collective variable into two independent
contributions that act at two well-separated time scales is
reasonable. Looking at the numerical values of the fitted parameters
in Table I one can can verify that $\tau_2 \sim t_{TOT} \gg \tau_1$.


\begin{figure}
  \includegraphics[width=0.95\columnwidth]{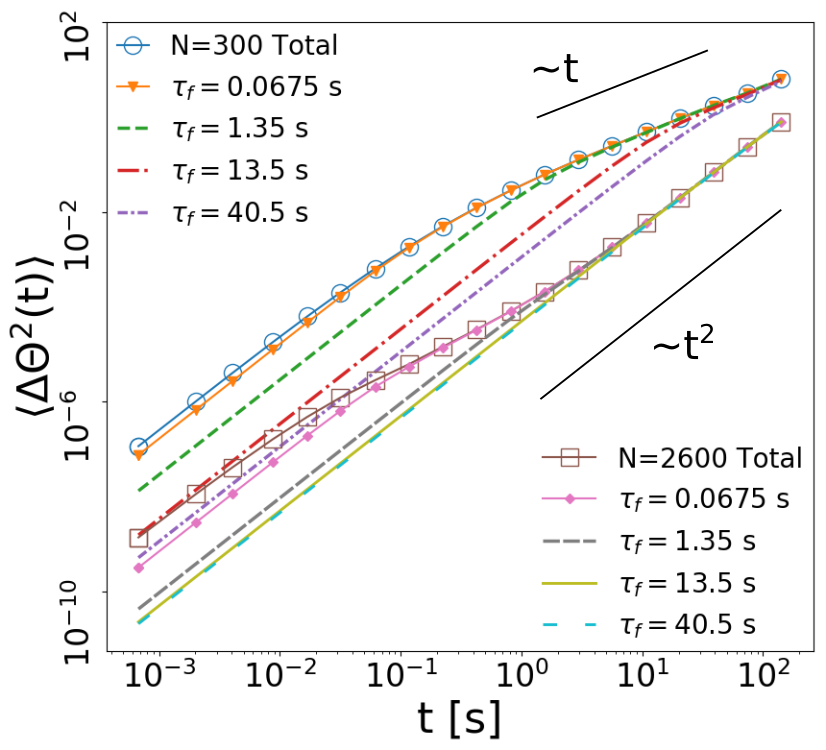}
	\caption{A: MSD of the total signals and the filtered ones for several $\tau_f$s and two values of $N$ with fixed $\Gamma=39.8$.  } \label{fig4:SlowAndTotal}
\end{figure}

In Fig. \ref{fig3:firstCompariWithData}b we show the MSD of $\Omega$
and $\Omega_s$ for two values of $\Gamma$, one cold case and one hot
case at similar density. For the cold case, i.e. at low $\Gamma$, we
see that $\langle \Delta \Theta(t)^2\rangle$ is ballistic at all times
except for the initial ones (the fast component is very weak so the
slow one emerges immediately). In the warmer case we can clearly
distinguish the two contributions to the total MSD: the fast one that
dominates the first times with its ballistic part and then diffuse,
and the slow one that dominates the late times with
superdiffusion. Regarding the filtered MSDs we can see that in both
cases it emerges in the total one exactly at the beginning of the
superdiffusive regime, at late times. In principle it is not obvious
that the filter used for $\Omega_s$ is able to isolate the slow
component of a sum of two signals. We can clearly expect that $\langle
\Delta \Theta_s(t)^2\rangle$ has to overlap the total MSD for $t >
\tau_f$ but how can we be sure that it is really describing the MSD of
$\Omega_2(t)$ i.e. the slow component of $\Omega$? We try to answer to
this question with Fig. \ref{fig4:SlowAndTotal} where we study the
effect of the filter on two qualitatively different signals for
several choices of $\tau_f$. We first discuss the MSD of the collective
variable in a dilute case (N=300) where we have an ordinary
ballistic-diffusive behavior with just one relevant time
scale~\cite{plati2019dynamical}. We can see that in this case the
filter lowers the energy of the ballistic part and stretches the
ballistic part up to times $t\sim\tau_f$ where the filtered MSD reunites
with the original one. In the dense case (e.g. low $\Gamma$) the MSD has the
ballistic-diffusive-superdiffusive behavior with two relevant time
scales (the fast $\tau_1$ and the slow $\tau_2$ of the aforementioned
model) but we can consider a third one $\tau^*$ defined as the time
for which the slow variable starts to dominate the total MSD (so the
time when the late superdiffusion starts, close to $\tau_2$).  In this
case the filter acts as in the previous one but for $\tau_f > \tau^*$ it
reaches a kind of saturation and leaves the shape of the MSD unchanged
(see the yellow and the cyan-dashed lines that overlap). This implies
that for $\tau_f > \tau^*$ the time in which $\langle \Delta
\Theta(t)^2\rangle$ and $\langle \Delta \Theta_s(t)^2\rangle$ reunite
coincide with $\tau^*$ and no more with $\tau_f$. In view of these last
analysis we can conclude that if in our data a contribution of a slow
variable is present, this filter operation tends to isolate it.

To sum up, in this section we have provided evidences of two main facts:

\begin{itemize}
\item The collective variable behaves as the sum of two
  independent OU processes with different characteristic times
  $\Omega(t)=\Omega_1(t)+\Omega_2(t)$;
  
\item A running average
  $\Omega_s(t)=\tau_f^{-1}\int_{t}^{t+\tau_f}\Omega(t')dt'$ with $\tau_f$
  larger than the time $\tau_1$ in which the fast component dominates,
  successfully isolates $\Omega_2(t)$ i.e. the slow component of
  $\Omega$.
  
\end{itemize}
\section{Motion of a rotating tracer (three time scales)}

Once we have a satisfying model for the collective granular motion, a
model for the tracer's motion can be studied on a solid basis,
with the aim of improving the model
in~\cite{lasanta2015itinerant}. The final model for the tracer appears
as a system of three equations for three variables (where actually two
of them are independent from each other).  To build this new model we
have considered that the tracer is moving in a complex granular fluid:
such a fluid has two characteristic time scales, as explained in the
previous section. The coupling between the tracer and the background
granular fluid can be modeled in two different ways: as a conservative
interaction that depends upon a relative {\em position} between the
tracer and a representative fluid particle, or as a viscous
interaction that depends upon the relative {\em velocity} between the
tracer and the fluid. The first choice was adopted
in~\cite{lasanta2015itinerant}, but our present study convinced us
that the second choice gives a better comparison with data. For a
viscous interaction, we were inspired by~\cite{sarra10b} were a linear
model for a massive granular tracer in a granular fluid (in a planar,
not cylindrical, geometry) was considered. The tracer - characterized
by velocity $V(t)$ - was coupled with the fluid - characterized by
local velocity $U(t)$ - through a viscous drag term proportional to $V(t)-U(t)$.
Depending on the specific region of the parameter space, this model
can reproduce both ordinary Brownian motion and cage effects with the
ballistic-caged-diffusive behavior in the MSD and the back-scattering
peak in the PDS. Therefore, it only lacks the superdiffusion at late
times to properly describe the phenomenon under interest here. As we
explained in section II, we cannot expect such a complex behavior from
a linear model with two variables, therefore we have to insert a third
one trying to be as less artificial as possible. To do so we leave the
equation for the tracer unchanged and complicate the expression of the
auxiliary variable making it coincide with the collective variable
$\Omega$ relative to the sub-set of the granular particles in the bulk
that actually  influence the tracer
dynamics. This quantity is always modeled by
Eqs. \eqref{eq:casiSlowFast1} and defined by Eq. \eqref{eq:OmegaDef1}
but with the mean operation extended just on the aforementioned
sub-set of grains. We expect for it the same \emph{qualitative}
behavior (but in general not quantitative) of the global collective
variable studied in the previous section: for this reason we decided not to
introduce a new symbol for it.

We end up with a three-variable
linear model defined by the following equations:
\begin{subequations} \label{eq:ModelPlati}
\begin{align}  
I\dot{\omega }(t)=-\gamma (\omega (t)-\Omega(t))+\sqrt{2T\gamma }\eta(t) \label{eq:ModelPlati1}\\
\Omega(t)=\Omega_1(t)+\Omega_2(t)\label{eq:ModelPlati2}\\
I_1\dot{\Omega}_1(t)=-\gamma_1\Omega_1(t) -\gamma_c\omega (t)+\sqrt{2T_1\gamma_1}\eta_1(t)\\
I_2\dot{\Omega}_2(t)=-\gamma_2\Omega_2(t)+\sqrt{2T_2\gamma_2}\eta_2(t)   .
\end{align}
\end{subequations}
Here we see that the bulk collective variable $\Omega$ acts as a local
velocity field on the tracer whose velocity is $\omega(t)$. At the same time, the fast
component of the collective variable $\Omega_1$ is coupled with $\omega$ through a viscous constant $\gamma_c$. These last two ingredients originate the observed broad cage effect.  The slow component
$\Omega_2$ is independent from the other variables and, as suggested
by the numerical analysis, it will emerge at late times in the MSD and
at small frequency in the PDS.  Regarding the physical meaning of the
coefficients we have for each variable an inertia $I_i$, a viscous
coefficient $\gamma_i$ and a temperature $T_i$.  We note that the
dilute limit (simple OU process for the tracer) is recovered by
sending $\gamma_1/I_1 \rightarrow \infty$ and $\gamma_2/I_2
\rightarrow \infty$ while the model for the collective variable alone
(Eqs. \ref{eq:casiSlowFast1}) is obtained by setting $\gamma_c=0$.  As
already noted in previous studies \cite{sarra10b,sarra12}, we recall
here that Eqs. \eqref{eq:ModelPlati} are equivalent to a Generalized
Langevin Equation with exponential memory which is consistent with a
typical approximation done for Brownian Motion when, at high
densities, the coupling of the tracer with fluid hydrodynamics modes,
decaying exponentially in time (see \cite{Z01}, Cap 8.6 and 9.1), must
be taken into account.

From Eqs. \eqref{eq:ModelPlati}, exploiting the formalism of
the multivariate linear stochastic processes~\cite{G90}, we can
compute (see Appendix) the PDS of the tracer.  In
Fig. \ref{fig5:ModelTracer} we show a comparison with the experimental and numerical
data, finding good agreement in all the frequency regimes.
\begin{figure}
  \includegraphics[width=\columnwidth]{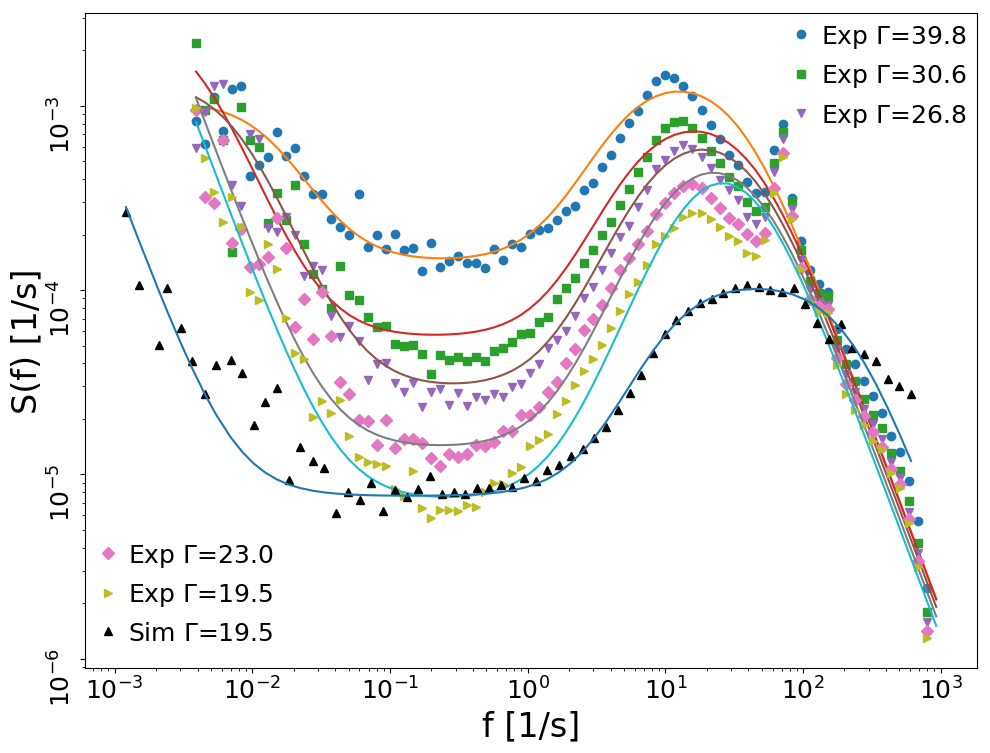}
	\caption{Comparison between experimental/numerical data and model
          (Eq.~\eqref{eq:ModelPlati}) for the PDS of the tracer for many values of $\Gamma$ and fixed $N=2600$. See caption of Fig. \ref{fig2} for explanation of peak close to $100$ Hz in the experimental spectrum.
        }
          \label{fig5:ModelTracer}
\end{figure}
The improvement with respect to the model defined by
Eqs. \eqref{lsmodel} and \eqref{model5} regards the form of the PDS. In the previous model,
the peak and the valley of the PDS are specular i.e. their position
and width depend upon the same combination of parameters so they
cannot be changed independently (Fig. \ref{fig2}). The experimental and
numerical PDS show instead that the valley and the back-scattering peak
are never specular and in general this is coherent with the scenario
suggested by the numerical simulations. Indeed, the valley is actually
the crossover between the motion of the tracer inside the cage and the
movement of the cage itself that enslaves the tracer. Once verified
the presence of a persistent collective motion of the granular medium
we can say that the cage moves as the collective variable on two time
scales that are, with a good approximation, independent. In this
picture it is thus reasonable that the frequency (time) where the slow
component emerge in the PDS (MSD) could change independently from what
is happening at the characteristic frequency (time) of the
back-scattering (cage) effect and vice-versa. This is possible with
Eqs. \eqref{eq:ModelPlati} by changing a combination of two parameters and leaving the others fixed
(Fig. \ref{fig6:ManyCurves}).


Both for Eqs. \eqref{eq:casiSlowFast1} and \eqref{eq:ModelPlati} the number of independent parameters needed is actually smaller than the one used but we have kept some redundancy for the purpose of presenting more clear model equations. Now, rescaling all the equations by the inertia $I_i$ we can rewrite the two models in a compact form that is more suitable for reading the next section. For the tracer we have:
\begin{subequations} \label{eq:ModelPlatiResc}
\begin{align}  
\dot{\omega }(t)=-\frac{1}{\tau} (\omega (t)-\Omega(t))+\sqrt{2\frac{q}{\tau} }\eta(t) \label{eq:ModelPlatiResc1}\\
\Omega(t)=\Omega_1(t)+\Omega_2(t)\\
\dot{\Omega}_1(t)=-\frac{1}{\tau_1}\left(\Omega_1(t) +\alpha\omega (t)\right)+\sqrt{2\frac{q_1}{\tau_1}}\eta_1(t) \label{eq:ModelPlatiResc2}\\
\dot{\Omega}_2(t)=-\frac{1}{\tau_2}\Omega_2(t)+\sqrt{2 \frac{q_2}{\tau_2}}\eta_2(t)   .
\end{align}
\end{subequations}
where $\tau_i=I_i/\gamma_i$, $q_i=T_i/I_i$ and
$\alpha=\gamma_c/\gamma_1$. Regarding the model for the collective variable
alone we note that the rescaled form of Eqs. \eqref{eq:casiSlowFast1}
(for the granular bulk dynamics when the tracer is absent) coincide
with Eqs. (\ref{eq:ModelPlatiResc}b-d) once set $\alpha=0$. 

The last aspect of the proposed model we want discuss regards the physical meaning of couplings between variables in system of Langevin equations. First we note that also without the introduction of ausiliary variables the effect on the tracer of the surrounding fluid is intrinsically contained in the Langevin approach. Indeed, also for ordinary Brownian motion (Eq. \eqref{eq:ModelPlatiResc1} without $\Omega(t)$) the characteristic time $\tau$ and the stationary variance $q$ depend on the properties of both tracer and fluid ~\cite{L08}. The need of additional variables emerges just in presence of multiple time scales in the tracer dynamics. In view of this, referring to Eqs. \ref{eq:ModelPlatiResc}, we can say that from an \emph{energetic} point of view the effect of the granular medium on the tracer can be contained in $q$ alone. Indeed, provided that $\tau \ll \tau_1$ and $q \gg q_1$, the fluctations in the steady states of $\omega$ are not affected by the introduction of $\Omega_1$. We can expect that this limit holds in our system because the inertia of the local collective variable $I_1$ is reasonably higher than the one of the blade $I$. We will confirm this expectation in the next section in which our fitting procedure shows that $q$ and the variance of $\omega$ are almost coincident as in a single variable process.

 These features are independent of $\alpha$ in Eq. \eqref{eq:ModelPlatiResc2} so, in this limit, $\alpha$ represents the adimensional strength of a coupling that affects just the memory and not the energy of the tracer dynamics. In other words, the introduction of $\Omega_1$ changes the shape of the PDS of $\omega$ leaving its integral unaltered (we remember here that if $\langle \omega \rangle=0$ then $\int_0^{\infty}df S(f)=\langle \omega ^2 \rangle$).  
Remarkably, studying the derivative of $S(f)$, it is possible to see that $\alpha>0$ is a necessary condition for the occurrence of the back-scattering effect. So, in our model, this effect is possible only if the tracer is coupled with a variable that is influenced by the tracer itself. This fact is compatible with the intuitive physical mechanism with which back-scattering is rationalized: the surrounding fluid is perturbed by the intruder and the latter feels with some delay in time the effect of this perturbation.
In view of this mutual influence, we find even more appropriate the definition of $\Omega$ as a local field in Eqs. \eqref{eq:ModelPlati} and \eqref{eq:ModelPlatiResc}. Indeed, even if the whole granular medium can be reasonably unperturbed by the intruder, there will be always a local fraction of it that reciprocally interacts with the tracer givig rise to the back-scattering effect.

This clarifies also the way in which the tracer is coupled with $\Omega_2$ that is not affected by $\omega$. Indeed, the cage of surrounding grains has two main effects: confining the tracer with back-scattering (coupling between $\omega$ and $\Omega_1$ with reciprocal influence) and dragging it into the slow dynamics (coupling between $\omega$ and $\Omega_2$ without reciprocal influence).

\begin{figure}
  \includegraphics[width=\columnwidth]{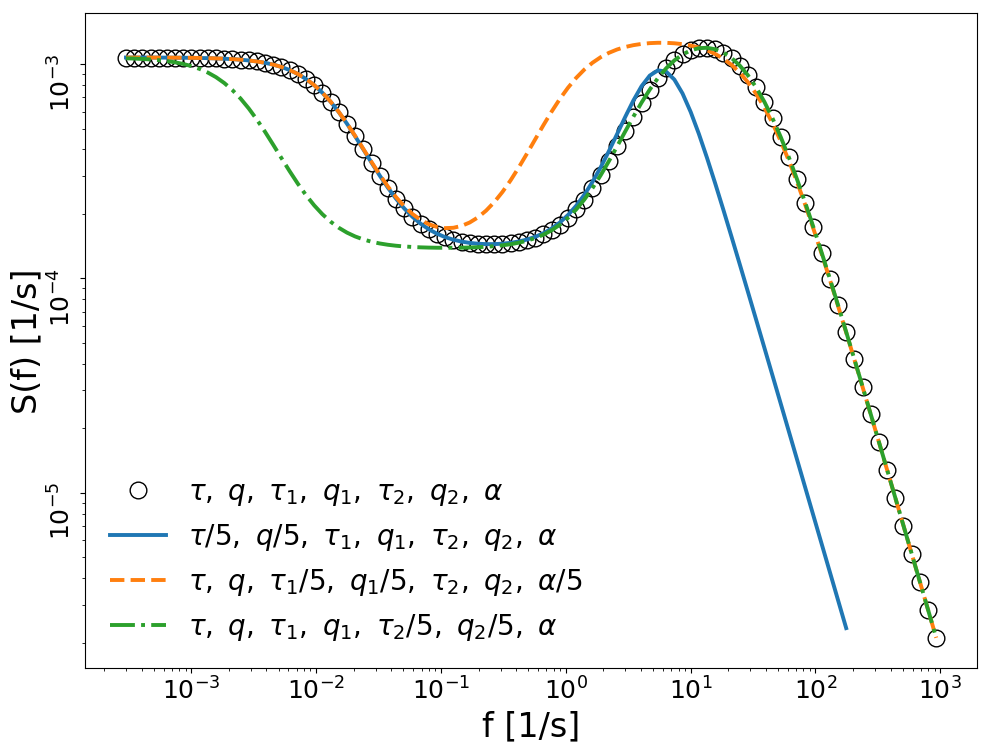}
	\caption{Four different shapes of $S(f)$, obtained by plotting
          the PDS of $\omega(t)$ computed from
          model~\eqref{eq:ModelPlati} or different arbitrary choices
          of the parameters where $\tau_i=I_i/\gamma_i$, $q_i=T_i/I_i$ and $\alpha=\gamma_c/I_1$. We demonstrate that it is possible to change the
          broadness of the valley independently from the one of the
          back-scattering peak and vice-versa.} \label{fig6:ManyCurves}
\end{figure}


\section{Physical meaning of the model's parameters}

In this Section we attempt to systematically fit our model's
parameters with the results of the numerical simulations and
experiments. This task has two main motivations. First, it may suggest
a way to infer or guess the model's parameters (or their behavior when
physical parameters are varied) in general situations. Second, it
makes more robust the identification of the model: a fuzzy or
unintelligible behavior of the model's parameter would be the symptom
of a weakness of the model itself.

\begin{figure}
  \includegraphics[width=0.49\columnwidth]{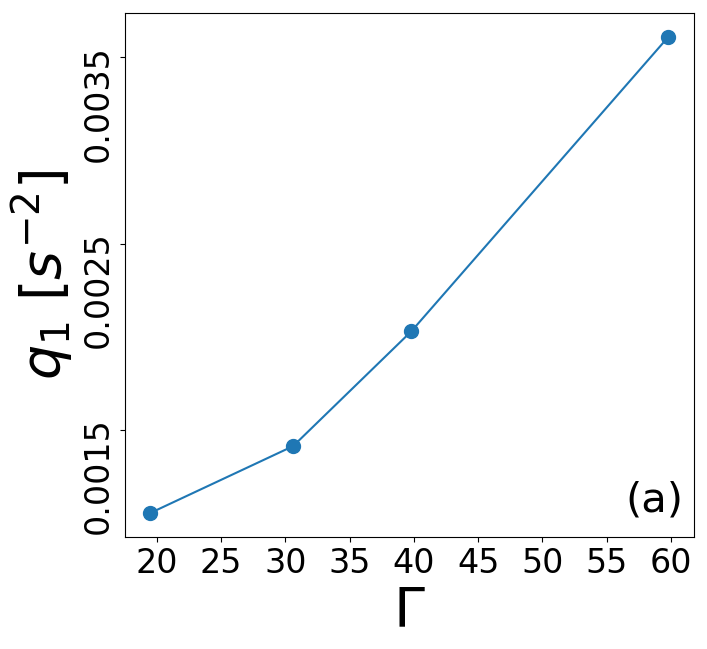}
\includegraphics[width=0.49\columnwidth]{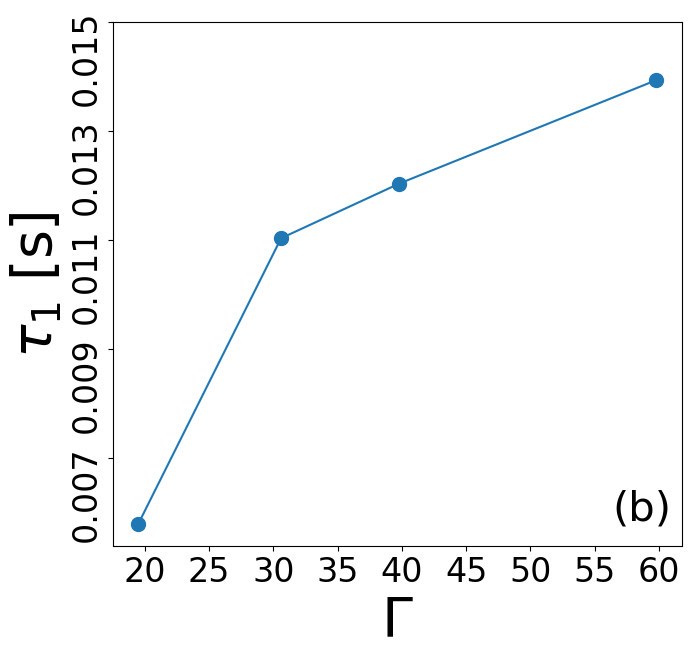}
    \includegraphics[width=0.49\columnwidth]{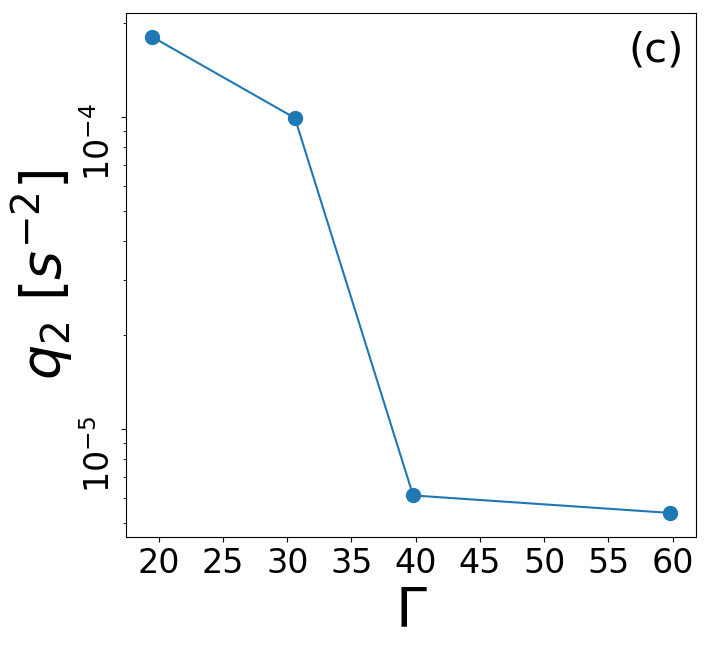}
  \includegraphics[width=0.49\columnwidth]{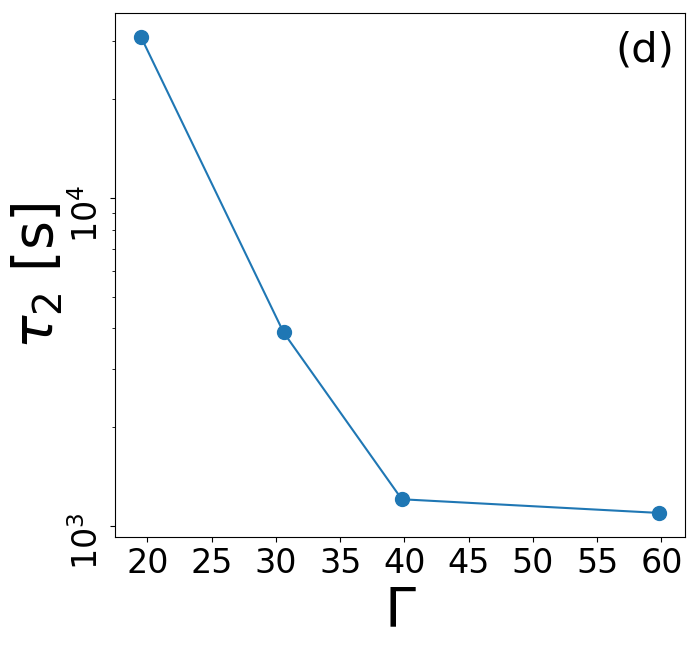}

	\caption{Fitted parameter
for the collective variable (simulations) versus $\Gamma$ with fixed $N=2600$. Fit has been done via PDS (Eq. \eqref{eq:casiSlowFast2pds}) relative to Eqs. (\ref{eq:ModelPlatiResc}b-d) with $\alpha=0$. Numerical values of the parameters are also reported in Table I.} \label{fig7:FitOfParamVarColl}
\end{figure}

For this purpose we report in Table I the fitted parameters obtained
for the collective variable (simulations) via
Eqs. (\ref{eq:ModelPlatiResc}b-d) with $\alpha=0$ and for the tracer
(experiments and simulations) via Eqs. \eqref{eq:ModelPlatiResc} for many values of
$\Gamma$. We first concentrate on the numerical data for the
collective variable $\Omega$. From Eqs. \eqref{eq:casiSlowFast1} and
\eqref{eq:ModelPlatiResc} it is clear that the sum of two independent
OU processes depends on four parameters: $\tau_1$, $q_1$, $\tau_2$,
$q_2$ where $\tau_i$ is the characteristic time of the single process
and $q_i$ is its variance.  In Section IV we verified that
Eqs. \eqref{eq:casiSlowFast1} properly reproduce the functional form
of the numerical PDS and MSD, now we want to study how the fitted
parameters behave as a function of $\Gamma$. In particular, we are
interested to verify if their numerical values reflect the physical
intuitions on which the model with two variables is based. In
Fig. \ref{fig7:FitOfParamVarColl} we can clearly see that $\tau_1$ and
$q_1$ are increasing with $\Gamma$ while $\tau_2$ and $q_2$ are
decreasing. The behavior of $q_1$ is intuitive because we can
reasonably think that this parameter grows with the ``temperature'' of
the physical external driving (the shaker). The behavior of $\tau_2$
and $q_2$ corroborates our intuition that reducing $\Gamma$ induces
the emergence of a slow time scale whose persistency time ($\tau_2$) and
intensity ($q_2$) grow. The microscopic origin of this fact can be
understood by considering the dynamical heterogeneity present in this
system (see for instance~\citep{plati2019dynamical}): local
temperature and pressure may vary a lot in space. When a collective
motion emerges, we find that a great fraction of particles
participates - as a condensed state - to the collective motion and a
smaller one still exhibits a gas-like behavior. We can actually
imagine that the slow contribution $\Omega_2$ to the total variable
$\Omega$ is mainly due to the particles in this condensed phase, whose
number could be thought as proportional to the effective inertia of
$\Omega_2$. At this point it is clear that increasing $\Gamma$ reduces
the fraction of particles in the condensed phase and consequently
reduces $\tau_2$.  The growth of $\tau_1$ with $\Gamma$ has not an
easy explanation in our opinion. In dilute kinetic models the
dissipative drag (here inverse of $\tau_1$) is often related to the
mean collision frequency (mediated through a ratio of masses and other
factors): however this quantity may have opposite trends when $\Gamma$
grows, i.e. it may increase because there is more energy (faster
collisions) or it may decrease because there is a larger mean free
path. Apparently the second phenomenon dominates the first. The
connection between dissipation and collision frequency, however, is
reasonable for dilute gases but certainly not obvious in condensed
phases.

\begin{table*}[]
\begin{tabular}{|c|c|c|c|c|c|c|c|c|}
\hline
& $\tau_1$ [s] & $q_1$ $[\text{s}^{-2}]$ & $\tau_2$ [s]& $q_2$ $[\text{s}^{-2}]$ & $\tau$ [s]& $q$ $[\text{s}^{-2}]$& $\alpha$ \\ \hline 
$\Gamma$=19.5 Tracer Simulations & 9.6e-02 & 5.7e-03 & 1.1e+04 & 2.0e-02 & 7.3e-04 & 4.4e-01 & 5.0 \\
$\Gamma$= 19.5 Coll. Var. Sim. with blade & 5.8e-03 & 1.1e-03 & 1.1e+04 & 8.1e-04 & 	\textbackslash & 	\textbackslash & 	\textbackslash \\
$\Gamma$= 19.5 Coll. Var. Simulations & 5.8e-03 & 1.1e-03 & 3.1e+04 & 1.8e-04 & 	\textbackslash & 	\textbackslash & 	\textbackslash \\
$\Gamma$= 30.6 Coll. Var. Simulations & 1.1e-02 & 1.4e-03 & 3.9e+03 & 9.9e-05 & 	\textbackslash & 	\textbackslash & 	\textbackslash \\
$\Gamma$= 39.8 Coll. Var. Simulations & 1.2e-02 & 2.0e-03 & 1.2e+03 & 6.1e-06 & 	\textbackslash & 	\textbackslash & 	\textbackslash \\
$\Gamma$= 59.8 Coll. Var. Simulations & 1.4e-02 & 3.6e-03 & 1.1e+03 & 5.4e-06 & 	\textbackslash & 	\textbackslash & 	\textbackslash \\ \hline

$\Gamma$=19.5 Tracer Experiments & 1.03e-01 & 1.0e-03 & 4.6e+03 & 4.2e-01 & 2.8e-03 & 4.5e-01 & 6.8 \\
$\Gamma$=23.0 Tracer Experiments & 1.06e-01 & 1.1e-03 & 7.0e+03 & 5.1e-01 & 2.8e-03 & 5.1e-01 & 5.0 \\
$\Gamma$=26.8 Tracer Experiments & 1.09e-01 & 1.2e-03 & 2.0e+01 & 4.7e-03 & 3.0e-03 & 6.2e-01 & 3.7 \\
$\Gamma$=30.6 Tracer Experiments & 1.10e-01 & 1.4e-03 & 4.0e+01 & 3.4e-03 & 3.3e-03 & 7.3e-01 & 2.8 \\
$\Gamma$=39.8 Tracer Experiments & 1.11e-01 & 2.0e-03 & 1.1e+01 & 2.5e-03 & 4.2e-03 & 9.5e-01 & 2.1 \\

\hline                  
\end{tabular}
\caption{Numerical values of the fitted parameters for the collective variable in simulations via Eqs. (\ref{eq:ModelPlati}b-d) with $\alpha=0$ and the tracer in experiments and simulations via Eqs. \eqref{eq:ModelPlatiResc}. The first and the second rows refer to the same simulation \emph{with} the blade. \label{tab:fitParam}}
\end{table*}

\begin{figure}
  \includegraphics[width=0.49\columnwidth]{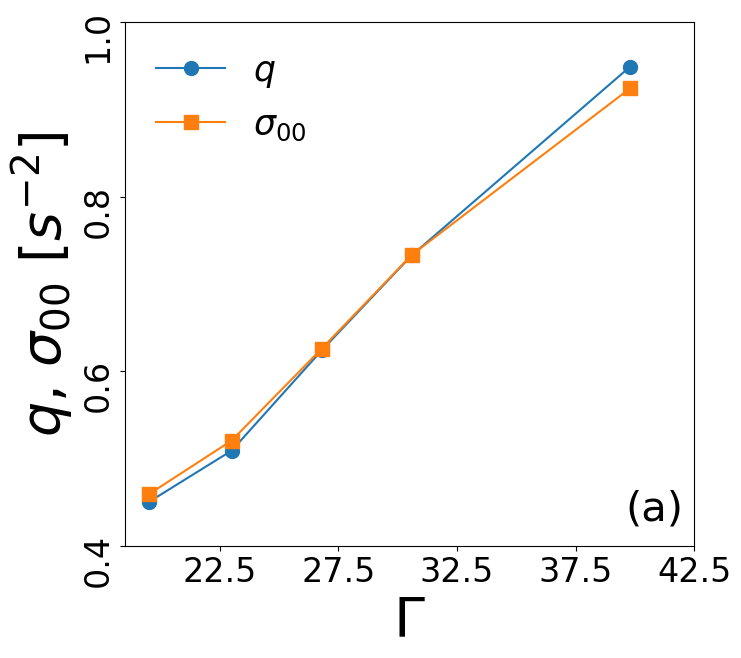}
    \includegraphics[width=0.49\columnwidth]{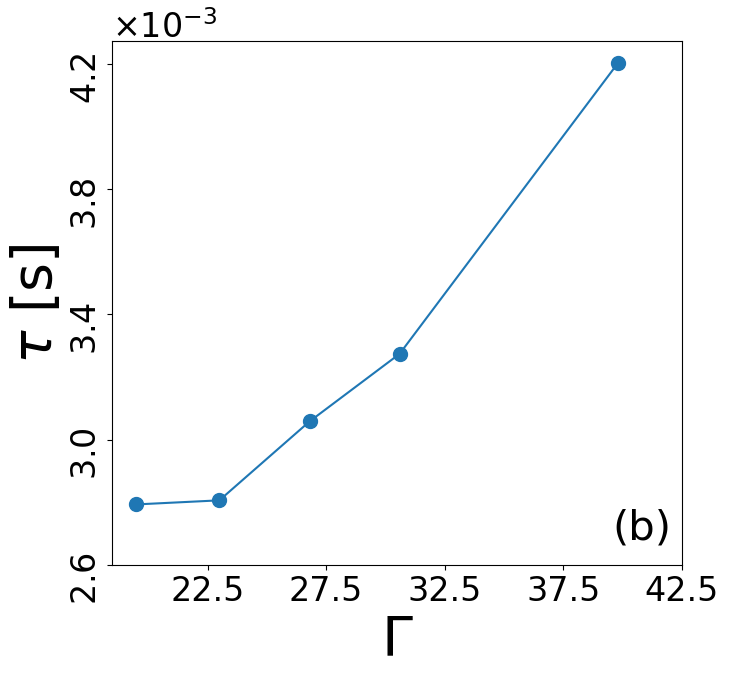}
\includegraphics[width=0.49\columnwidth]{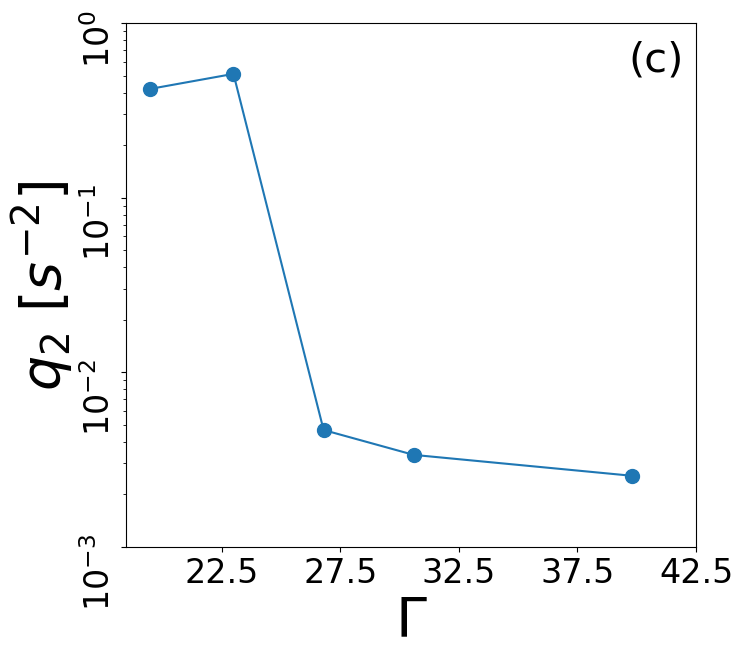}
  \includegraphics[width=0.49\columnwidth]{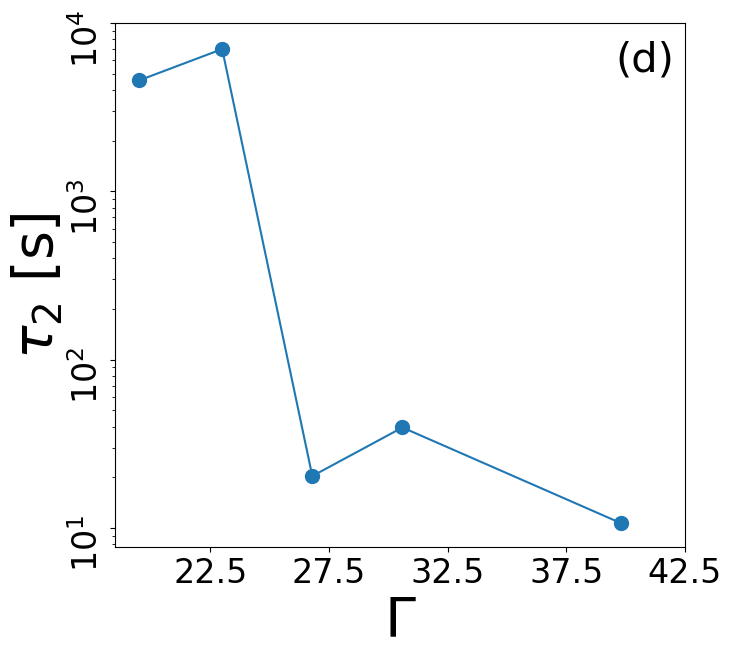}

	\caption{Fitted parameters for experimental tracer versus $\Gamma$ with fixed $N=2600$.  Fit has been done via PDS (see Eq. (A7) in the Appendix) relative to
Eqs. \eqref{eq:ModelPlatiResc}. Numerical values of the parameters are also reported in Table I. } \label{fig8:FitOfParamTracer}
\end{figure}

Regarding the experimental data of the tracer, from
Eqs. \eqref{eq:ModelPlatiResc} we see that the number of free
parameters for a fit is 7 and they are
$\tau,q,\tau_1,q_1,\tau_2,q_2,\alpha$. We recall that $q$, $q_1$ and
$q_2$ would be equal to the stationary variances of $\omega$,
$\Omega_1$ and $\Omega_2$, respectively, if these variables were not
coupled. In our model with couplings one must consider the covariance
matrix $\sigma$ (see Appendix A.2) which is related to the noise
amplitudes through a relation which also involves the coupling
matrix~\cite{G90}. However, in view of the discussion about couplings at the end of Sec. V, and with the aim of reducing the freedom in the
fitting procedure, we decided to set $q$ to coincide with the variance
of $\omega$ measured in experiments. We then verify \emph{a
  posteriori} how g
  ood is this approximation. In
Fig. \ref{fig8:FitOfParamTracer}a we compare experimental $\langle
(\omega - \langle\omega \rangle) ^{2} \rangle $ with the theoretical
one calculated with fitted parameters and verify that our assumption
is reasonable.  The behavior of $q$ together with the ones of $\tau$,
$\tau_2$ and $q_2$ shown in the same figure (panels b, c and d), is
coherent with the phenomenology already explained for fast and slow
parts of the collective variable of Fig. \ref{fig7:FitOfParamVarColl}
(we remind that $\Omega_2$ is totally independent from the other
variables so $q_2$ always coincides with the variance of $\Omega_2$).
Indeed, the idea of our model is to consider the tracer ($\omega$)
enslaved by the collective variable: at short times it feels the
effect of the fast component $\Omega_1$ (high frequency decay and
back-scattering peak) while at late times $\Omega_2$ starts to
dominate the entire dynamics with its persistent ballistic drifts. The
motion of the tracer is then characterized by more than one
characteristic time scale.  Looking at Eq. \eqref{eq:ModelPlati1} it
is quite natural to associate $\tau$ and $q$ respectively to the
characteristic time and the variance of the short-time dynamics of
$\omega$. It is therefore reassuring to find for these parameters
similar behaviors to those observed for $\tau_1$ and $q_1$ in
numerical simulations (compare panels a and b of
Fig. \ref{fig7:FitOfParamVarColl} and \ref{fig8:FitOfParamTracer}). Also the values of $\tau_1$ and $q_1$ obtained from the experimental data via Eqs. \eqref{eq:ModelPlatiResc} (not shown in figures but reported in Table I) follow the same qualitative behavior.  
Regarding couplings, our fitting procedure revealed the situation depicted at the end of Sec. V. Looking at Table I and Fig. \ref{fig8:FitOfParamTracer} we find that the effect on the tracer of the auxiliary variables is negligible form an \emph{energetic} point of view ($q \simeq (\omega - \langle\omega \rangle) ^{2} \rangle $) but not from a \emph{memory} one  ($\alpha \sim \mathcal{O}(1-10)$ for all the fitted spectra).  

The results reported in Fig. \ref{fig8:FitOfParamTracer} regard the
experimental tracer but, as is shown for one case ($\Gamma=19.5$) in
Fig. \ref{fig5:ModelTracer}, we have also tested our model on the
numerical data coming from the simulations \emph{with} the blade. We
note that the good agreement shown is obtained for values of
$\tau_1$, $q_1$, $\tau_2$ and $q_2$ that are quite different from the
ones coming from a direct fit of the two components of the collective
variable in simulations \emph{with} or \emph{without} the blade (see
first, second and seventh rows of Table I). This is not surprising
because, as already stated at the end of section V, the variable
$\Omega$ actually coupled to the tracer, in Eqs.\eqref{eq:ModelPlati}
and \eqref{eq:ModelPlatiResc}, is a \emph{local} collective
variable. The latter reasonably differs - quantitatively - from the
global one because of a considerable spatial heterogeneity of granular
temperature and diffusivity \cite{plati2019dynamical}.


\section{Conclusions}

In conclusion, we have proposed a series of linear stochastic models
to rationalise a series of experimental and numerical results. In the
phenomenon we have tackled, vibrofluidized dense granular materials
display persistent slow drifts superimposed to fast collisional
processes. We stress that the proposed models are purely
phenomenological. They can be considered analogous to hydrodynamic
models for dense fluids (even granular fluids), where the transport
coefficients are not derived from microscopic parameters but are
obtained from empirical observations.  An important added value of
phenomenological models for slow variables is to offer arguments in
favor of the idea of scale separations, which is not always guaranteed
in granular fluids~\cite{G99,kadanoff99,baldovin19}.

We have built two main models. The one for the motion of the angular
drift of the granular medium, which is a sum of two independent
Langevin diffusions (i.e. Brownian motion with inertia),
characterizing fast and slow scales: the independence is consistent
with the lack of cage effects which are usually an evidence of
coupling between fast and slow scales. The presence of inertia also in
the slow mode provides the main ingredient for persistent ballistic
motion. The second model is for the motion of a rotating tracer
immersed in the granular medium, which is the most accessible variable
in experiments. Such a model is built upon the idea that the tracer is
coupled with a local fraction of the surrounding medium through a purely
viscous interaction, similar to other granular tracer models. Since
both models contain several independent parameters, it is not surprising
that they reproduce all available data. Less trivial is the fact
that the fitted parameters behave in a coherent way and are consistent
with physical intuition, as shown in Section VI. We have also solved
an important inconsistency of a previous model for the
tracer~\cite{lasanta2015itinerant}, which was unable to describe the
power spectrum at medium time-scales.

We hope to stimulate future theoretical investigations in order to
derive these models, or part of their parameters, following a kinetic
theoretical approach.

\begin{acknowledgments}
The authors are indebted to Marco Baldovin for fruitful scientific
discussions. The authors acknowledge the financial support of Regione
Lazio through the Grant "Progetti Gruppi di Ricerca" N. 85-2017-15257
and from the MIUR PRIN 2017 project 201798CZLJ.
\end{acknowledgments}


\appendix

\section{Multivariate Linear Stochastic Processes}

\subsection{Generic process with two variables}
A generic multivariate linear stochastic process can be written as $\dot{X}(t)=-AX(t)+B\tilde{\eta}(t)$, where $X(t)$ is a vector of variables, $\tilde{\eta}(t)$ a vector of  white noises with zero mean and unitary variance while $A$ and $B$ are the matrices that define characteristic times, diffusion coefficients and eventually couplings of the variables. In two dimension, assuming independent noises between the variables, we have $X(t)=\{x_0(t),x_1(t)\}$ and:  
\begin{equation}
A=\left(\begin{matrix}a & b\\c & d\end{matrix}\right), \quad B=\left(\begin{matrix}D_1 & 0\\0 & D_2\end{matrix}\right), \quad
\tilde{\eta}(t)=\left(\begin{matrix}\eta_1(t) \\ \eta_2(t)\end{matrix}\right).
\end{equation}
We can now compute the spectral matrix through the following relation \citep{G90}: 
\begin{equation}
S(f)=\dfrac{1}{2\pi}(A+i2\pi f)^{-1}BB^T(A^T-i2\pi f)^{-1},
\end{equation}\label{eq:specRel}obtaining for the spectrum of a single variable (for example $x_0(t)$):
\begin{equation}
S_{00}(f)=\frac{\frac{1}{2\pi} D_{1}^{2} d^{2} + 2 \pi D_{1}^{2} f^{2} + \frac{1}{2\pi} D_{2}^{2} b^{2}}{ \left(16 \pi^{4} f^{4} + 4 \pi^{2}f^{2} \left( a^{2} + 2 b c + d^{2}\right)+a^{2} d^{2} - 2 a b c d + b^{2} c^{2} \right)}.
\end{equation}
Computing the first derivative we find $S_{00}'(f)=P(f)/Q(f)$ where $P(f)=c_1f^5+c2f^3+c3f$ with $c_1,c_2<0$. This polynomial cannot have a double stationary point in the region $f>0$ so there is not any choice of parameters that reproduce the behavior under interest here i.e. a low-frequency decay followed by a back-scattering peak. These results does not change if we take $x_1(t)$ instead of $x_0(t)$.

\vspace{0.1cm}

\subsection{Model for The Tracer}
From Eqs. \eqref{eq:ModelPlati}  we have $\dot{X}(t)=-AX(t)+B\tilde{\eta}(t)$ with $X(t)=\{\omega (t),\Omega_1(t),\Omega_2(t)\}$ and:
\begin{eqnarray}\label{matrdefini}
&&A=\left( {\begin{array}{ccc}
    \mu & -\mu & -\mu \\
    \mu_1 & \alpha  & 0\\
   0& 0 & \mu_2 \\
  \end{array} } \right), \\
&&B=\left( {\begin{array}{ccc}
    \sqrt{2\mu q} & 0 & 0\\
   0 & \sqrt{2 \mu_1 q_1}  & 0\\
   0& 0 & \sqrt{2\mu_2 q_2} \\
  \end{array} } \right), \\
&&\tilde{\eta}=\left( {\begin{array}{c}
    \eta (t)\\
   \eta_1(t)\\
   \eta_2(t)\\
  \end{array} } \right).
\end{eqnarray}
where $\mu_i=1/\tau_i=\gamma_i/I_i$, $q_i=T_i/I_i$ and $\alpha=\gamma_c/\gamma_1$. The covariance matrix $\sigma$ is realted to $A$ and $B$ through the relation: $A\sigma + \sigma A^T=B B^T$. We have computed the spectrum used for the fit of sec. V and VI always through Eq. (A2) obtaining Eq. (A7) where $\hat{f}=2\pi f$:

\begin{widetext}
\begin{equation}
S_{00}(\hat{f})=\frac{\mu/\pi \left[ \hat{f}^4q + \hat{f}^2\left( \mu_1^{2} q + \mu_1 \mu q_{1} + \mu_2^{2} q + \mu_2 \mu q_{2}\right) + \mu_1^{2} \mu_2^{2} q + \mu_1^{2} \mu_2 \mu q_{2} + \mu_1 \mu_2^{2} \mu q_{1}\right]}{ \hat{f}^{6} + \hat{f}^4\left(\mu_1^{2} +  \mu_2^{2}  +  \mu^{2} - 2 \mu \alpha \right)   + \hat{f}^2\left( \mu_1^{2} \mu_2^{2}  +  \mu_1^{2} \mu^{2} + 2 \mu_1 \mu^{2} \alpha +  \mu_2^{2} \mu^{2} - 2 \mu_2^{2} \mu \alpha +\mu^{2} \alpha^{2}\right)  + \mu_2^2 \mu^2(\alpha+\mu_1)^2} .
\end{equation} \label{eq:specLong}
\end{widetext}


\bibliographystyle{apsrev4-1}
\bibliography{biblio}
\end{document}